\documentclass[aps,prl,twocolumn,showpacs]{revtex4}
\usepackage{graphicx}
\usepackage{dcolumn}
\usepackage{amsmath}

\begin{document}

\title{Parametric excitation of a Bose-Einstein condensate 
       in a 1D optical lattice} 

\author{M.~Kr\"amer, C. Tozzo and F. Dalfovo }
\affiliation{INFM-BEC and Dipartimento di Fisica, Universit\`a di Trento, 
I-38050 Povo, Italy }

\date{February 10th, 2004}

\begin{abstract}
We study the response of a Bose-Einstein condensate to a periodic
modulation of the depth of an optical lattice. Using Gross-Pitaevskii
theory, we show that a modulation at frequency $\Omega$ drives the 
parametric excitation of Bogoliubov modes with frequency $\Omega/2$.  
The ensuing nonlinear dynamics leads to a rapid broadening of the 
momentum distribution and a consequent large increase of the condensate 
size after free expansion. We show that this process does not require 
the presence of a large condensate depletion. Our results reproduce 
the main features of the spectrum measured in the superfluid phase 
by St\"oferle {\it et al.}, Phys. Rev. Lett. {\bf 92}, 130403 (2004). 
\end{abstract}


\maketitle

Exploring the effects of interaction and low dimensionality is one 
of the central themes in current research on ultracold Bose gases. 
Optical lattices are ideal tools in this sense. A major achievement 
was the demonstration that, by increasing the depth of an optical 
lattice, a quantum phase transition occurs from the superfluid to 
a Mott insulating phase \cite{greiner}. More recently, St\"oferle 
{\it et al.} \cite{stoeferle,schori} made an effort to characterize 
this phenomenon by measuring the excitation spectrum. The atoms are 
loaded into the ground state of an harmonic trap plus an optical 
lattice, and they are subsequently excited by modulating the lattice 
depth. Then the gas is released from the trap and the width of the 
expanding cloud is taken as a measure of the excitation energy. The 
quantum phase transition shows up as the crossover from a broad 
continuum of excitations in the superfluid phase to a discrete 
spectrum in the Mott insulating phase. 

The broad resonance in the superfluid regime was unexpected.  In 
fact, the Gross-Pitaevskii (GP) theory predicts that, in the linear 
response regime, a modulation of the lattice depth cannot excite the 
gas \cite{menotti}. For this reason, it was conjectured that 
the measured spectra might be the effect of strong interactions, 
not included in the GP theory, which yield a significant 
quantum depletion of the condensate \cite{stoeferle,blatter}.
Here we show that the observed resonance is caused by the 
parametric amplification of Bogoliubov states and the subsequent 
nonlinear dynamics, which leads to the broadening of the momentum 
distribution. We describe both processes by using the GP theory 
and assuming the initial condensate to be out of equilibrium. A 
tiny initial occupation of Bogoliubov modes is found to be enough 
to produce a strong response at high frequency, while larger 
fluctuations are needed to explain the observed spectrum at low 
frequency. 

We simulate the dynamics of the condensate by solving the GP 
equation \cite{rmp} for an axially symmetric condensate of 
$^{87}$Rb, similar to typical ``tubes" in the experiments of 
Ref.~\cite{stoeferle}. Since the transverse and axial trapping 
frequencies are such that $\omega_\perp/\omega_z \gg1$, one can 
factorize the order parameter in the product of a Gaussian radial 
component of $z$-and $t$-dependent width, $\sigma(z,t)$, and an 
axial wave function  $\Psi(z,t)$. The GP equation yields 
\cite{npse} 
\begin{equation}
i\hbar\frac{\partial\Psi}{\partial t} = 
\left[ -\frac{\hbar^2}{2m} \left( \partial_z^2
+\frac{1}{\sigma^2} +\frac{\sigma^2}{a_\perp^4} \right)
+ V +\frac{gN|\Psi|^2}{2\pi\sigma^2} \right] \Psi \, ,
\label{npse}
\end{equation}
and $\sigma=a_{\perp}(1+2aN|\Psi|^2)^{1/4}$. Here
$a_{\perp}=[\hbar/(m\omega_{\perp})]^{1/2}$, $N$ is the number of
particles, $g=4\pi\hbar^2 a/m$, with $a$ the s-wave scattering 
length and $m$ the atomic mass.  The axial potential $V$ is the 
sum of the harmonic trap and the optical lattice:
\begin{equation}
V (z,t) = (m/2) \omega_z^2 z^2 
        + s E_R [1 + A \sin(\Omega t)]
         \sin^2 (q_B z)  \,,
\label{v}
\end{equation}
where $q_B=\pi/d$ is the Bragg wave vector, $d$ is the lattice 
spacing and $s$ is the lattice depth in units of the recoil energy
$E_R=\hbar^2 q_B^2/2m$.  To simulate a single tube of 
Ref.~\cite{stoeferle}, we use $d=413$ nm, $\omega_z=2\pi \times 
84.6$~Hz and $\omega_{\perp}=2 \pi \times 36.5$~KHz, yielding 
$E_R/\hbar = 2\pi \times 3.34$ KHz. We first 
calculate the ground state for given $N$ and $s$, by solving the
stationary GP equation with $A=0$. Then we solve the
time-dependent equation (\ref{npse}) with $A \neq 0$ for
a modulation time $t_{\rm m}$. Finally, we switch off the 
external potential and we let the condensate expand for $25$ ms 
\cite{note0}. 

\begin{figure}[h]
\begin{center}
\includegraphics[width=8cm]{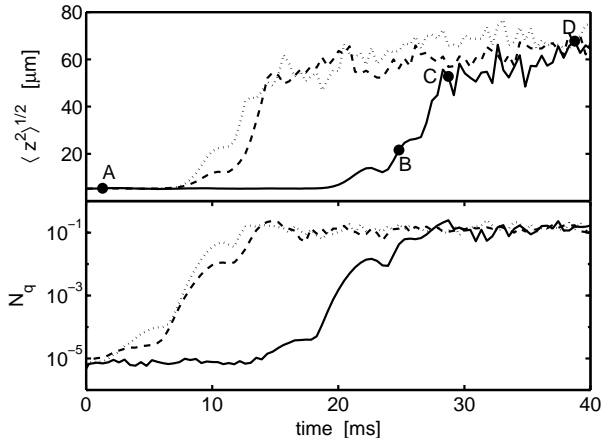}
\end{center}
\caption{Upper panel: the width of the density distribution $\langle
z^2\rangle^{1/2}$ after $25$ ms of free expansion is plotted as a 
function of the modulation time $t_{\rm m}$, as obtained with 
$\Omega= 1.33 E_R/\hbar$ and $A=0.2$; Lower panel: 
population $N_q$ of the parametrically excited mode 
with $q=0.56 q_B$ and $\omega=\Omega/2$. The different lines 
refer to different initial states. Solid line: ground state; 
dashed: ground state plus a weak Bragg pulse; dotted: ground 
state plus a small white noise. }
\label{fig1}
\end{figure}

In Figs.~1 and 2 we report the results for $N=100$, $s=4$, 
$A=0.2$ and $\Omega=1.33 E_R/\hbar$. The solid line 
in the upper panel of Fig. \ref{fig1} shows the width of the density 
distribution at the end of the expansion as a function of the 
in-trap modulation time $t_{\rm m}$ \cite{note1}. Before about 
$20$ ms nothing seems to occur. Then the width starts increasing 
and eventually reaches values of about $10$ times the size of 
the expanding ground state. The density and momentum distributions 
of the in-trap condensate at the instants A, B, C, and D are 
plotted in Fig.~2. At the early time A, both are still 
indistinguishable from those of the ground state. In contrast, 
the momentum distribution in B clearly features new sharp peaks 
at $\pm q$, with $q=0.56 q_B$ and the density is modulated on a 
lengthscale of about $4$ sites, accordingly.  By the time 
C, the populations at $\pm q$ have grown large. Besides, the 
lateral peaks at $\pm 2q_B$, associated with the density modulation 
of the ground state in the lattice, are smaller and contributions 
from momenta throughout the first and second Brillouin zones start 
showing up. At D, the system hardly bears any resemblance with 
the ground state and its broad momentum distribution explains the 
increase of the width of the expanding cloud shown in 
Fig.~\ref{fig1}. 

\begin{figure}[h]
\begin{center}
\includegraphics[width=8.5cm]{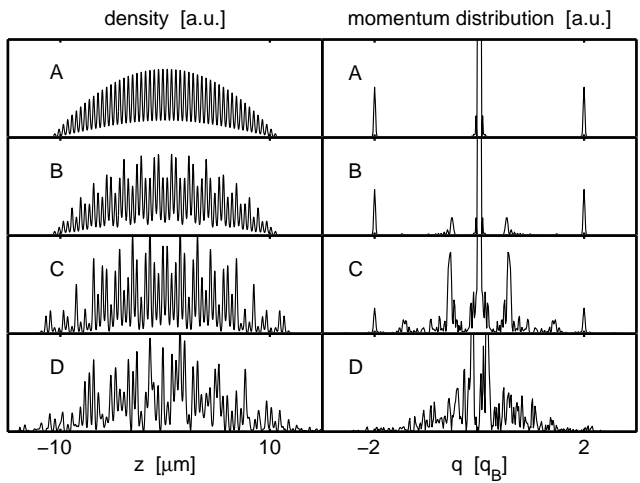}
\end{center}
\caption{Density (left) and momentum (right) distributions at 
different times A, B, C and D given in Fig.~\ref{fig1}. }
\label{fig2}
\end{figure}

This behavior is a consequence of the parametric instability of
Bogoliubov modes. A parametric instability corresponds to the
exponential growth of certain modes of a system induced by the
periodic variation of a parameter \cite{landau}. Systems governed by a
nonlinear Schr\"odinger equation, as the GP equation, can exhibit a
wide variety of parametric instabilities (see e.g., \cite{rapti} and
references therein). In our case, the modulation of the lattice depth
at frequency $\Omega$ causes the instability of modes with frequency 
$\omega(q) = \Omega/2$. To demonstrate this, we perform simulations 
at various $\Omega$ and, from each one, we extract the positions 
$\pm q$ of the first extra peaks which grow in the momentum
distribution. Given these values of $q$, in Fig.~3 we plot the
frequencies $\Omega/2$ as a function of $q$ (points), together
with the dispersion law of the lowest Bogoliubov band, $\omega(q)$
(solid line). The latter is calculated by keeping the same optical
lattice and transverse trap, but neglecting the weak axial
confinement; the number of atoms per lattice site is taken to be 
equal to the average number of atoms per site in the actual 
trapped condensate \cite{note2,epjd}.  The agreement between points 
and solid line clearly indicates that the dynamics of the condensate 
is directly related to the spectrum of Bogoliubov modes.

\begin{figure}[h]
\begin{center}
\includegraphics[width=7cm]{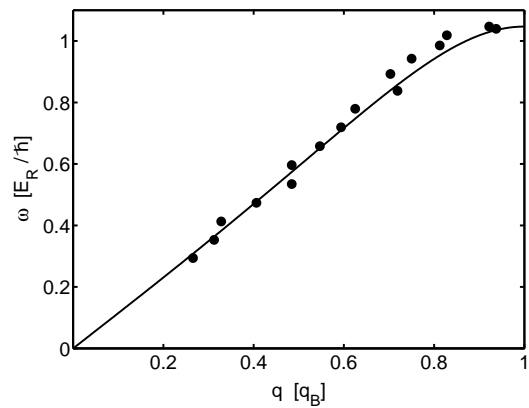}
\end{center}
\caption{For each modulation of given frequency $\Omega$,
we draw a point at $(q,\omega)$, where $q$ is
the wavevector of the parametrically excited mode and 
$\omega=\Omega/2$. 
The solid line is the lowest Bogoliubov band. }
\label{fig3}
\end{figure}

In order to be parametrically amplified, the modes at $\pm q$ 
must preexist at $t=0$. This means that our initial 
state is not a pure ground state. Indeed some numerical noise is 
always present and yields a very small occupation of Bogoliubov 
modes. To understand this point, we define the fraction of 
atoms, $N_q(t)$, which contributes to the peaks at $\pm q$ 
\cite{definition} and plot it in Fig.~1. The solid line starts
at about $5 \times 10^{-6}$ which, for the parameters of this 
simulation, coincides with the $\pm q$ components of the exact 
ground state. These nonvanishing components are due to the 
finite axial size of the condensate; they do not grow in time and 
must not be confused with the population of the Bogoliubov modes. 
The seed population is instead provided by the numerical noise 
and is much less than $10^{-6}$ at $t=0$. Then it grows exponentially 
and it becomes visible in Fig.~1 after about $15$ ms, when it 
exceeds the ground state level. It eventually saturates at a value 
of the order of $10^{-1}$. The time scale of this growth is much 
longer than the lattice modulation period. The growth rate that 
we extract from our simulations is linear in $\Omega$ and $A$ within
our accuracy.  This linear dependence agrees 
with the results of a semi-analytic model which describes the 
mean-field coupling between the ground state and $\pm q$ 
excitations within a quasiparticle projection method \cite{next}.  

The time needed to saturate depends on the amount of seed. To 
explore this dependence, we mimic an initial 
depletion by introducing some extra ``noise" in a controllable 
way. For instance, just before $t=0$ we excite the condensate 
with a short and weak Bragg pulse in resonance with the Bogoliubov 
mode at $q$.  Alternatively, we simply add random noise with 
uniform (white noise) or thermal-like distributions. The effects 
of the Bragg pulse and white noise 
are shown in Fig.~\ref{fig1}, where the seed is chosen to be 
$\sim 10^{-5}$ in both cases. The main effect of this extra seed 
is to anticipate the instant at which the parametric excitation 
becomes noticeable. Yet, the growth rate and the level at 
which $\langle z^2\rangle^{1/2}$ and $N_q$ saturate remain 
the same as for the solid line, which corresponds to less than 
$10^{-15}$ seed excitations. This means that, for this $\Omega$, 
the amount of seed which is enough to trigger the parametric 
instability is indeed very small. In the experiments the system is 
certainly not in its groundstate at $t=0$ due to thermal and 
quantum depletion, and to excitations produced when loading the 
condensate into the lattice \cite{note4}. Our calculations show 
that even a very tiny deviation from the groundstate can 
produce a strong response on the timescale of the experiment.

The evolution leading up to the results in Figs.~1 and 2 not 
only involves the parametric excitation of the $\pm q$ modes, but 
also the subsequent broadening of the momentum distribution, as 
in C and D. This is due to the nonlinear mean-field dynamics 
of the condensate governed by the GP equation. This nonlinear 
dynamics is triggered by the growth of the parametrically unstable 
modes, but its outcome is little affected by changing $\Omega$ 
or $A$. Even more, it is hardly influenced by the external 
perturbation; in fact, switching off the lattice modulation at time 
C has very little effect on the outcome at time D or later. 
This implies that the response of the condensate as a function 
of $\Omega$ mainly reflects the $\Omega$-dependence 
of the growth rates of the parametrically excited modes. Moreover,
the time scale for the nonlinear broadening of the momentum 
distribution is also relatively long, so that the increase 
of the width $\langle z^2\rangle^{1/2}$ occurs on a scale of 
several tens of ms, as in Fig.~1. This agrees with the behavior 
observed in the experiments (see Fig.~3 of Ref.~\cite{koehl}).    

As seen in Fig.~3, the main effect of changing the lattice 
modulation frequency is that of picking out a different Bogoliubov 
mode for parametric excitation. In the limit $\Omega\to 0$ the 
growth rate vanishes and the width $\langle z^2\rangle^{1/2}$
at a fixed $t_{\rm m}$ becomes equal to the width of a condensate
which expands without being modulated. The same limiting value is 
obtained on the opposite side, namely at $\Omega$ close to twice 
the Bogoliubov bandwidth, $2\omega(q_B)$, where parametric 
excitations are not possible due to the presence of the gap in 
the Bogoliubov spectrum at the zone boundary. Therefore the 
response of the system takes the form of a broad peak in the 
range $0 < \Omega < 2 \omega(q_B)$. 

\begin{figure}[h]
\begin{center}
\includegraphics[width=8cm]{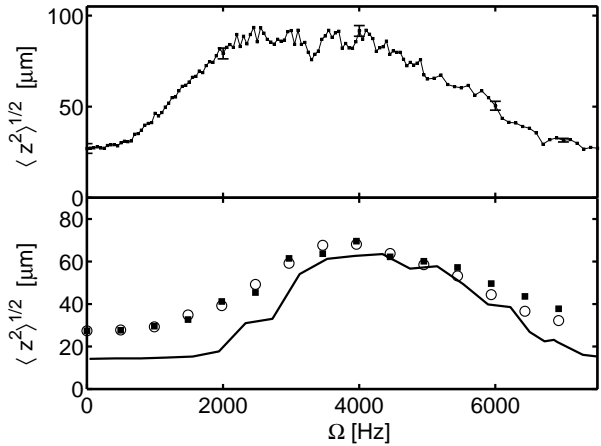}
\end{center}
\caption{Width $\langle z^2\rangle^{1/2}$ of the expanded 
array of condensates as function of $\Omega$. Top panel: 
experimental data of Ref.~\protect\cite{stoeferle}, for $s=4$; 
Solid line in bottom panel: our simulations starting from a 
small white noise. Empty circles and solid squares: same but 
with a much larger white noise and thermal-like noise, 
respectively (see text for explanation).}
\label{fig4}
\end{figure}

In order to compare our results with the experimental observations
of Ref.~\cite{stoeferle}, a simulation with a single tube is not 
enough. The actual condensate is made of an array of tubes having 
different population, $N$, and hence different Bogoliubov 
bandwidths. Given the total number of atoms in the array and the 
trapping frequencies, one can estimate the population of each tube
by calculating the ``average" density distribution within a 
local density approximation as in Refs.~\cite{epjd,meret}. Then 
we perform several simulations for different values of $N$ and we
sum their contributions to $\langle z^2\rangle^{1/2}$ using 
the ``average" density distribution as a weighting function. The 
central condensate is taken to have $N=90$ which is expected to be 
reasonably close to reality. The results are shown in the lower 
panel of Fig.~4. The solid line is the result of GP simulations
with a white noise as a seed excitation, corresponding to 
$N_q \sim 10^{-3}$ at $t=0$. This small seed is enough to produce
a broad response which qualitatively agrees with the experimental
one (upper panel). The height of the peak is of the same order and 
its slope at high frequency is very similar. Within our approach, 
this slope originates from the fact that the response vanishes at 
twice the Bogoliubov bandwidth, $2 \omega(q_B)$, but this bandwidth 
is different for tubes with different $N$ in the array and this 
provides a smoothing of the high frequency tail. By repeating the 
calculation for a deeper lattice ($s=6$, instead of $4$), we find 
that this tail shifts to lower frequencies by about $10$-$15$\%. 
This agrees with the experimental observations and is consistent 
with the calculated decrease of the bandwidth. 

On the other side of the peak, at low frequency, the response turns
out to be more sensitive to the seed. In order to reproduce the
observed response around $2000$ Hz, we have to start with a condensate
in which a significant number of atoms populate the lowest Bogoliubov
modes. This also yields an increase of the axial width of the expanded
condensate without modulation. The effect of a larger seed is
represented in Fig.~4 by empty circles and solid squares, which are
the results of calculations with a white noise and a thermal-like
seed, respectively.  The thermal seed is simulated by random
fluctuations, where the average occupation number of Bogoliubov
excitations follows a Bose distribution at a given temperature. The
dispersion $\omega(q)$ of a uniform condensate, calculated as for
Fig.~3, and the corresponding quasiparticle amplitudes are used to
relate the occupation of the quasiparticle states to the momentum
distribution of the atoms. Both the temperature in the thermal-like
seed and the level of the white noise are chosen in such a way that
the width without modulation is close to the experimental value,
$\langle z^2\rangle^{1/2} \simeq 25 \mu$m. This implies a total
depletion of the order of $20$\% in the case of white noise and the
temperature $k_B T \sim 2 \times 10^{-3} \hbar \omega_\perp$ for 
thermal noise. The latter value is compatible with the experimental
estimate $k_B T < 6 \times 10^{-3} \hbar \omega_\perp$ given in
Ref.~\cite{moritz}. As one can see, in both cases the overall response
is higher and broader, and the qualitative agreement with the
experiments is improved. This result is also compatible with the
observation made in \cite{schori} that the excitation efficiency
decreases when the transverse confinement is reduced and, hence, the
initial depletion of the condensate is reduced as well.  A
quantitative comparison is however difficult at this level. In the
experiment a thermalization process is performed after the modulation
and before the expansion, by lowering the depth of the transverse
optical lattice and letting the tubes interact for a while. This is
not included in our simulations, since it would require a full 3D
calculation. Also the procedure of averaging over many tubes, some of
them with only a few atoms, limits the accuracy of the comparison.
Nevertheless, our analysis suggests that a characterization of
classical and/or quantum fluctuations can indeed be possible by
appropriately choosing the trapping geometry and the experimental
procedure and using the parametric resonances for a selective
amplification of initial fluctuations \cite{next}.

In conclusion we have shown that the broad resonance observed 
in condensates subject to a periodic modulation of the optical 
lattice can be interpreted as the effect of a parametric 
instability of Bogoliubov modes. The occurrence of this type of 
instability in Bose-Einstein condensates is an interesting 
property of these systems, which originates from their 
superfluid nature and the existence of undamped collective 
excitations. The parametric instability and the subsequent nonlinear 
dynamics are nicely accounted for by the GP equation. Since our 
calculations are purely mean-field in nature, they are not conclusive 
in excluding possible many-body processes in the explanation of the 
experimental observations. Exploiting all predictions 
of GP theory is however essential in view of investigations on 
beyond mean-field effects due to strong interactions and low 
dimensionality. 

\acknowledgments 
We thank T.~Esslinger and his group for stimulating discussions
and for sharing their data.  Discussions with L.~Pitaevskii, 
I.~Carusotto and H.P.~B\"uchler are gratefully acknowledged.


\begin{thebibliography}{20}
\bibitem{note1} As in the experiments, when calculating the width 
$\langle z^2 \rangle^{1/2}$ we ignore the lateral peaks due to the
$\pm 2q_B$ components of the momentum distribution.
\bibitem{note2} As far as longitudinal excitations are concerned, 
the effects of the axial harmonic confinement on $\omega(q)$
are small and the condensate can be safely assumed to be 
infinite, thus allowing one to exploit the Bloch symmetry 
of Bogoliubov excitations.
\bibitem{definition} $N_q$ is the integral of the momentum 
distribution in an interval $\Delta q \sim 2\pi /Z$ around 
$\pm q$, divided by the total number of particles, where 
$Z$ is the axial size of the condensate.
\bibitem{note4}   A process in which $\pm q$ quasiparticles are 
created via stimulated emission in the linear response regime and in 
the presence of quantum depletion is suggested in 
Ref.~\protect\cite{blatter}. This mechanism can also provide a seed 
for the subsequent parametric amplification.
\end{thebibliography}
\end{document}